\documentclass[english]{IEEEtran}
\usepackage[T1]{fontenc}
\usepackage[latin9]{inputenc}
\usepackage{amsmath}
\usepackage{amsthm}
\usepackage{amssymb}
\usepackage{graphicx}

\makeatletter
\theoremstyle{definition}
\newtheorem{assumption}{Assumption}
\theoremstyle{plain}
\newtheorem{lem}{\protect\lemmaname}
\theoremstyle{plain}
\newtheorem{prop}{\protect\propositionname}
\theoremstyle{plain}
\newtheorem{thm}{\protect\theoremname}

\usepackage{cite}
\usepackage{multirow}

\makeatother

\usepackage{babel}
\providecommand{\lemmaname}{Lemma}
\providecommand{\propositionname}{Proposition}
\providecommand{\theoremname}{Theorem}

\begin{document}
\title{Distributed RISE-based Control for Exponential Heterogeneous Multi-Agent
Target Tracking of Second-Order Nonlinear Systems}
\author{Cristian F. Nino, Omkar Sudhir Patil, Sage C. Edwards, and Warren
E. Dixon\thanks{Cristian F. Nino, Omkar Sudhir Patil, Sage C. Edwards, and Warren
E. Dixon are with the Department of Mechanical and Aerospace Engineering,
University of Florida, Gainesville, FL, 32611, USA. Email:\{cristian1928,
patilomkarsudhir, sageeedwards, wdixon\}@ufl.edu.}\thanks{This research is supported in part by AFRL grant FA8651-24-1-0018
and AFOSR grant FA9550-19-1-0169. Any opinions, findings and conclusions
or recommendations expressed in this material are those of the author(s)
and do not necessarily reflect the views of the sponsoring agency.}}
\maketitle
\begin{abstract}
A distributed implementation of a Robust Integral of the Sign of the
Error (RISE) controller is developed for multi-agent target tracking
problems with exponential convergence guarantees. Previous RISE-based
approaches for multi-agent systems required 2-hop communication, limiting
practical applicability. New insights from a Lyapunov-based design-analysis
approach are used to eliminate the need for multi-hop communication
required in previous literature, while yielding exponential target
tracking. The new insights include the development of a new P-function
that works in tandem with the graph interaction matrix in the Lyapunov
function. Nonsmooth Lyapunov-based stability analysis methods are
used to yield semi-global exponential convergence to the target agent
state despite the presence of bounded disturbances with bounded derivatives.
The resulting outcome is a controller that achieves exponential target
tracking with only local information exchange between neighboring
agents.
\end{abstract}


\section{\label{sec:Introduction}Introduction}

\IEEEPARstart{M}{ulti-agent} systems have emerged as a powerful framework
for distributed sensing, decision-making, and control across various
domains. A challenging problem within this framework is target tracking,
where networked agents must collectively track a designated target
despite uncertainties and limited information exchange \cite{Lewis.Zhang.ea2013,Ren.Cao2013}.
This capability enables critical applications in surveillance, search
and rescue, environmental monitoring, and defense systems \cite{Kwon.Hwang2019,Mao.Jafarnejadsani.ea2020,Drew2021}.

The multi-agent target tracking problem presents several technical
challenges, including network topology constraints, communication
limitations, and the presence of unmodeled dynamics and external disturbances.
Robust control techniques that effectively handle these uncertainties
while maintaining stability guarantees are therefore essential. Among
these techniques, the Robust Integral of the Sign of the Error (RISE)
methodology has demonstrated significant potential by achieving exponential
tracking error convergence with continuous control inputs despite
time-varying disturbances \cite{Patil.Isaly.ea2022,Patil.Stubbs.ea2022,Patil.Kamalapurkar.ea2025}.

The RISE control approach incorporates integration of a signum term
to achieve robust performance, making it particularly suitable for
multi-agent systems where disturbance rejection is critical. Central
to the RISE control structure is the design of the $P$-function,
which is used in the Lyapunov-based stability analysis. The $P$-function
construction leverages convolutional integral properties and specific
1-norm identities. However, extending the traditional $P$-function
construction used in previous literature to multi-agent systems presents
significant challenges, because the stability analysis is typically
conducted at the ensemble level rather than the agent level.

The first application of RISE to multi-agent tracking was presented
in \cite{Klotz.Kan.ea2015}, where a synchronization error is used
to represent the collective tracking error, which includes an interaction
matrix to combine the graph Laplacian and pinning matrix. The controller
in \cite{Klotz.Kan.ea2015} requires 2-hop communication, meaning
agents need information from the neighbors of their neighboring agents,
which may limit some practical scenarios. RISE control has been applied
to related multi-agent problems, including the leader-following flocking
problem \cite{Wang.Sun.ea2022}, where a virtual leader guides the
multi-agent system, thereby avoiding challenges inherent to target
tracking where targets may be indifferent or noncooperative. Additionally,
\cite{Hu.Wen.ea2022} employed RISE control for distributed stabilization
in multi-agent systems with uncertain strong-weak competition networks,
addressing a fundamentally different problem from target tracking.

A distributed implementation of a RISE-based controller for multi-agent
target tracking is developed in this paper. The key innovation is
the inclusion of the graph interaction matrix directly in the Lyapunov
function, rather than coupling it explicitly to the signum term as
in \cite{Klotz.Kan.ea2015}. This modification eliminates the need
for 2-hop communication but introduces new challenges in the $P$-function
construction. To overcome these challenges, a novel $P$-function
design is developed that incorporates the interaction matrix while
enabling the use of 1-norm identities. Through a nonsmooth stability
analysis, semi-global exponential convergence to the target agent
state is established.

The primary contributions of this paper are: (1) a distributed RISE-based
control design that requires only local information exchange, (2)
a novel $P$-function construction technique for multi-agent systems
with interaction matrix incorporation, and (3) exponential convergence
despite the presence of disturbances and uncertainties.

\section{Preliminaries}

\subsection{Notation}

For a set $A$ and an input $x$, the indicator function is denoted
by $\boldsymbol{1}_{A}\left(x\right)$, where $\boldsymbol{1}_{A}\left(x\right)=1$
if $x\in A$, and $\boldsymbol{1}_{A}\left(x\right)=0$ otherwise.
For a vector $\mathbf{x}\in\mathbb{R}^{n}$, the element-wise signum
function is given by $\text{sgn}\left(\mathbf{x}\right)=\left[\begin{array}{ccc}
\text{sgn}\left(x_{1}\right) & \cdots & \text{sgn}\left(x_{n}\right)\end{array}\right]^{\top}\in\mathbb{R}^{n}$, where each $x_{i}\in\mathbb{R}$ and $\text{sgn}\left(x_{i}\right)=\mathbf{1}_{\left\{ x>0\right\} }\left(x_{i}\right)-\mathbf{1}_{\left\{ x<0\right\} }\left(x_{i}\right)$.
Let $\mathbf{1}_{n}\in\mathbb{R}^{n}$ denote the column vector of
length $n>1$ whose entries are all ones. Similarly, $\mathbf{0}_{n}\in\mathbb{R}^{n}$
denotes the column vector of length $n>1$ whose entries are all zeroes.
For $m,n>1$, $\mathbf{0}_{m\times n}\in\mathbb{R}^{m\times n}$ denotes
the $m\times n$ zero matrix. The $p\times p$ identity matrix is
denoted by $I_{p}$. Given $M\in\mathbb{Z}_{>0}$, define $\left[M\right]\triangleq\left\{ 1,2,\ldots,M\right\} $.
Let $n,m\in\mathbb{Z}_{>0}$ with $m>n$. The $p$-norm of $r\in\mathbb{R}^{n}$
is denoted by $\left\Vert r\right\Vert _{p}$, where in the special
case that $p=2$, $\left\Vert \cdot\right\Vert $ is used. The Euclidean
norm of $r\in\mathbb{R}^{n}$ is $\left\Vert r\right\Vert \triangleq\sqrt{r^{\top}r}$.
Given a positive integer $N$ and collection $\left\{ x_{i}\right\} _{i\in\left[N\right]}\subset\mathbb{R}^{n}$,
let $\left(x_{i}\right)_{i\in\left[n\right]}\triangleq\left[\begin{array}{ccc}
x_{1}^{\top} & \cdots & x_{N}^{\top}\end{array}\right]^{\top}\in\mathbb{R}^{nN}$. Let $f:\mathbb{R}^{n}\to\mathbb{R}$ be a differentiable function.
The maximum and minimum eigenvalues of $A=A^{\top}$ are denoted by
$\overline{\lambda}_{A}\in\mathbb{R}$ and $\underline{\lambda}_{A}\in\mathbb{R}$,
respectively.

\subsection{Nonsmooth Analysis}

The space of essentially bounded Lebesgue measurable functions is
denoted by $\mathcal{L}_{\infty}$. A function is of class ${\tt C}^{k}$
if it has $k$ continuous derivatives. The Lebesgue measure on $\mathbb{R}^{n}$
is denoted by $\mu$. Given some sets $A$ and $B$, a set-valued
map $F$ from $A$ to subsets of $B$ is denoted by $F:A\rightrightarrows B$.
The notation $\overline{\mathrm{co}}A$ denotes the closed convex
hull of the set $A$. The notation $\mathbb{B}_{\delta}\left(x\right)$,
for $x\in\mathbb{R}^{n}$ and $\delta>0$, is used to denote the set
$\left\{ y\in\mathbb{R}^{n}:\left\Vert x-y\right\Vert <\delta\right\} $.
The interior of a set $S$, denoted by $\text{int}\left(S\right)$,
is the set of all points $x\in S$ for which there exists $\delta>0$
such that $\mathbb{B}_{\delta}\left(x\right)\subset S$. The closure
of a set $S$, denoted by $\overline{S}$, is the union of $S$ and
its boundary points. A set is compact if it is closed and bounded,
and a set is precompact if its closure is compact. Consider a Lebesgue
measurable and locally essentially bounded function $h:\mathbb{R}^{n}\times\mathbb{R}_{\geq0}\to\mathbb{R}^{n}$.
The Filippov regularization of $h$ is defined as $K\left[h\right]\left(y,t\right)\triangleq\bigcap\limits _{\delta>0}\bigcap\limits _{\mu\mathbb{S}=0}\overline{\text{co}}\ h\left(\mathbb{B}_{\delta}\left(y\right)\backslash\mathbb{S},t\right)$,
where $\underset{\mu\mathbb{S}=0}{\bigcap}$ denotes the intersection
over all sets $\mathbb{S}$ of Lebesgue measure zero \cite[Equation 2b]{Paden.Sastry1987}.
Additionally, given any sets $A,B\subset\mathbb{R}$, the notation
$A\leq B$ is used to state $a\leq b$ for all $a\in A$ and $b\in B$.
The notation \textquotedbl a.e.\textquotedbl{} (almost everywhere)
means that a property $P$ holds for all $x\in X\backslash N$, where
$N\in\Sigma$ and $\mu\left(N\right)=0$ in a measure space $\left(X,\Sigma,\mu\right)$.
A function $y:\mathcal{I}_{y}\to\mathbb{R}^{n}$ is called a Filippov
solution of $\dot{y}=h\left(y,t\right)$ on the interval $\mathcal{I}_{y}\subseteq\mathbb{R}_{\geq0}$,
if $y$ is absolutely continuous on $\mathcal{I}_{y}$, and is a solution
to the differential inclusion $\dot{y}\stackrel{{\rm a.e.}}{\in}K\left[h\right]\left(y,t\right)$.
Clarke's generalized gradient for a locally Lipschitz function $V:\mathbb{R}^{n}\times\mathbb{R}_{\geq0}\to\mathbb{R}$
is defined as $\partial V\left(x,t\right)\triangleq\overline{\text{co}}\left\{ \lim\nabla V\left(x,t\right):\left(x_{i},t_{i}\right)\to\left(x,t\right),\left(x_{i},t_{i}\right)\not\in\Omega_{V}\right\} $,
where $\Omega_{V}$ denotes the set of measure zero wherever $\nabla V$
is not defined \cite[Def. 2.2]{Shevitz.Paden1994}.

\subsection{Algebraic Graph Theory}

Consider a network of $N$ distinct vertices indexed by $i\in\mathcal{V}$,
where $\mathcal{V}\triangleq\left[N\right]$ for $N\in\mathbb{Z}_{>0}$.
The flow of information between vertices is modeled by a static, weighted,
and undirected graph $G\triangleq\left(\mathcal{V},E,w\right)$, where
$E\subseteq\mathcal{V}\times\mathcal{V}$ represents the set of edges
between distinct vertices, and $w:E\to\mathbb{R}_{>0}$ is a function
that assigns a positive real weight $w\left(e\right)$ to each edge
$e\in E$. A bidirectional exchange of information between vertices
$i$ and $j$, denoted by $i\leftrightarrow j$, exists if and only
if $\left(i,j\right)\in E$ and $\left(j,i\right)\in E$ with $i\neq j$.
The neighborhood of vertex $i$, denoted by $\mathcal{N}_{i}$, is
the set of vertices $j\in\mathcal{V}$ such that information flows
between $i$ and $j$, i.e., $\mathcal{N}_{i}\triangleq\left\{ j\in\mathcal{V}:\left(j,i\right)\in E\wedge\left(i,j\right)\in E,j\neq i\right\} $.
The adjacency matrix $A\in\mathbb{R}^{N\times N}$ of the graph $G$
is defined by $a_{ij}\triangleq w\left(\left(j,i\right)\right)\mathbf{1}_{\left\{ \left(j,i\right)\in E\wedge\left(i,j\right)\in E\right\} }$,
the degree matrix $D\in\mathbb{R}^{N\times N}$ is defined by $D\triangleq\text{diag}\left\{ A\mathbf{1}_{N}\right\} $,
and the graph Laplacian $L\in\mathbb{R}^{N\times N}$ is defined by
$L\triangleq D-A$.

Consider an external vertex indexed by $0$, and define the extended
vertex set as $\overline{\mathcal{V}}\triangleq\mathcal{V}\cup\left\{ 0\right\} $.
The extended flow of information is modeled by a static graph $\overline{G}\triangleq\left(\overline{\mathcal{V}},\overline{E}\right)$,
where $\overline{E}\subseteq\overline{\mathcal{V}}\times\overline{\mathcal{V}}$
is the extended edge set. Exclusive to vertex $0$, the notation $0\to i$
indicates that vertex $i$ can sense vertex $0$, i.e., $0\to i$
exists if and only if $\left(0,i\right)\in\overline{E}$. If $0\to i$,
vertex $i$ is said to be pinned.

\section{Problem Formulation}

\subsection{System Dynamics}

Consider a multi-agent system composed of $N$ agents indexed by $\mathcal{V}$,
and consider a target agent indexed by $\left\{ 0\right\} $. The
dynamics for agent $i\in\mathcal{V}$ are given by\footnote{For addressing an uncertain $g_{i}$, potential extensions might leverage
techniques found in \cite{Teo.How.ea2009,MacKunis.Patre.ea2010}.}
\begin{equation}
\ddot{q}_{i}=f_{i}\left(q_{i},\dot{q}_{i},t\right)+g_{i}\left(q_{i},\dot{q}_{i},t\right)u_{i}\left(t\right)+d_{i}\left(t\right),\label{eq:Agent Dynamics}
\end{equation}
where: $q_{i},\dot{q}_{i},\ddot{q}_{i}\in\mathbb{R}^{n}$ denote the
agents' unknown generalized position, velocity, and acceleration,
respectively; the unknown functions $f_{i}:\mathbb{R}^{n}\times\mathbb{R}^{n}\times\left[0,\infty\right)\to\mathbb{R}^{n}$
and $d_{i}:\left[0,\infty\right)\to\mathbb{R}^{n}$ represent modeling
uncertainties and exogenous disturbances, respectively; $g_{i}:\mathbb{R}^{n}\times\mathbb{R}^{n}\times\left[0,\infty\right)\to\mathbb{R}^{n\times m_{i}}$
denotes a known control effectiveness matrix; and $u_{i}:\left[0,\infty\right)\to\mathbb{R}^{m_{i}}$
denotes the control input. Here, the functions $f_{i}$ are of class
$\mathtt{C}^{3}$, the mappings $t\mapsto f_{i}\left(q_{i},\dot{q}_{i},t\right)$,
$t\mapsto\nabla f_{i}\left(q_{i},\dot{q}_{i},t\right)$, and $t\mapsto\nabla^{2}f_{i}\left(q_{i},\dot{q}_{i},t\right)$
are uniformly bounded, the functions $g_{i}$ have full row-rank,
are of class $\mathtt{C}^{3}$, the mappings $t\mapsto g_{i}\left(q_{i},\dot{q}_{i},t\right)$,
$t\mapsto\nabla g_{i}\left(q_{i},\dot{q}_{i},t\right)$, and $t\mapsto\nabla^{2}g_{i}\left(q_{i},\dot{q}_{i},t\right)$
are uniformly bounded, and the functions $d_{i}$ are of class $\mathtt{C}^{2}$,
and there exist known constants $\overline{d}_{i},\overline{\dot{d}}_{i},\overline{\ddot{d}}_{i}\in\mathbb{R}_{>0}$
such that $\left\Vert d_{i}\left(t\right)\right\Vert \leq\overline{d}_{i}$,
$\left\Vert \dot{d}_{i}\left(t\right)\right\Vert \leq\overline{\dot{d}}_{i}$,
and $\left\Vert \ddot{d}_{i}\left(t\right)\right\Vert \leq\overline{\ddot{d}}_{i}$
for all $t\in\left[0,\infty\right)$, for all $i\in\mathcal{V}$.
By the full row-rank property, the existence of the right Moore-Penrose
inverse $g_{i}^{+}:\mathbb{R}^{n}\times\mathbb{R}^{n}\times\left[0,\infty\right)\to\mathbb{R}^{m_{i}\times n}$
is ensured, for all $i\in\mathcal{V}$, where the functions $g_{i}^{+}$
are of class $\mathtt{C}^{1}$, and the mappings $t\mapsto g_{i}^{+}\left(q_{i},\dot{q}_{i},t\right)$
and $t\mapsto\nabla g_{i}^{+}\left(q_{i},\dot{q}_{i},t\right)$ are
uniformly bounded for all $i\in\mathcal{V}$.

The dynamics for the target agent are given by
\begin{equation}
\ddot{q}_{0}=f_{0}\left(q_{0},\dot{q}_{0},t\right),\label{eq:Target Dynamics}
\end{equation}
where $q_{0},\dot{q}_{0},\ddot{q}_{0}\in\mathbb{R}^{n}$ denote the
target's unknown generalized position, velocity, and acceleration,
respectively, and the function $f_{0}:\mathbb{R}^{n}\times\mathbb{R}^{n}\times\left[0,\infty\right)\to\mathbb{R}^{n}$
is unknown, of class $\mathtt{C}^{3}$, and the mappings $t\mapsto\nabla f_{0}\left(q_{0},\dot{q}_{0},t\right)$
and $t\mapsto\nabla^{2}f_{0}\left(q_{0},\dot{q}_{0},t\right)$ are
uniformly bounded.
\begin{assumption}
\label{targetbounds}There exist known constants $\overline{q}_{0},\overline{\dot{q}}_{0},\overline{\ddot{q}}_{0},\overline{\dddot{q}}_{0}\in\mathbb{R}_{>0}$
such that $\left\Vert q_{0}\left(t\right)\right\Vert \leq\overline{q}_{0}$,
$\left\Vert \dot{q}_{0}\left(t\right)\right\Vert \leq\overline{\dot{q}}_{0}$,
$\left\Vert \ddot{q}_{0}\left(t\right)\right\Vert \leq\overline{\ddot{q}}_{0}$,
and $\left\Vert \dddot{q}_{0}\left(t\right)\right\Vert \leq\overline{\dddot{q}}_{0}$
for all $t\in\left[0,\infty\right)$.
\end{assumption}

\subsection{Control Objective}

Each agent $i\in\mathcal{V}$ can measure the relative position $q_{i,j}\in\mathbb{R}^{n}$
and relative velocity $\dot{q}_{i,j}\in\mathbb{R}^{n}$ between itself
and any agent $j\in\overline{\mathcal{N}}_{i}$ , where the relative
position is defined as
\begin{equation}
q_{ij}\triangleq q_{j}-q_{i}.\label{eq:Relative Position}
\end{equation}

The objective is to design a distributed controller that enables all
agents to track the target using only relative state measurements,
given the unknown dynamics described by (\ref{eq:Agent Dynamics})
and (\ref{eq:Target Dynamics}). To quantify the tracking performance,
the target tracking error $e_{i}\in\mathbb{R}^{n}$ is defined as
\begin{equation}
e_{i}\triangleq q_{0}-q_{i}.\label{eq: tracking error}
\end{equation}
Using (\ref{eq: tracking error}), the neighborhood tracking error
$\eta_{i}\in\mathbb{R}^{n}$ is defined as
\begin{equation}
\eta_{i}\triangleq b_{i}e_{i}+\sum_{j\in\mathcal{N}_{i}}a_{ij}q_{ij},\label{eq: n tracking}
\end{equation}
where $b_{i}\in\left\{ 0,1\right\} $ indicates whether agent $i\in\mathcal{V}$
senses the target. Using (\ref{eq:Relative Position}), (\ref{eq: n tracking})
is expressed in an equivalent analytical form as
\begin{equation}
\eta_{i}=b_{i}e_{i}-\sum_{j\in\mathcal{N}_{i}}a_{ij}\left(e_{j}-e_{i}\right).\label{eq:relative position error 2}
\end{equation}

\section{Control Design}

Let $\mathcal{I}\triangleq\left[t_{0},t_{1}\right]$ denote the interval
where solutions exist for the subsequent closed-loop error system,
where $t_{0},t_{1}\in\mathbb{R}_{\geq0}$ and $t_{1}>t_{0}$. Based
on the subsequent stability analysis, the continuous control input
$u_{i}$ for each agent $i\in\mathcal{V}$ is designed as
\begin{align}
u_{i} & =g_{i}^{+}\left(q_{i},\dot{q}_{i},t\right)\left(k_{3}\dot{\eta}_{i}+\left(\left(k_{1}+k_{2}\right)k_{3}+1\right)\eta_{i}\right.\nonumber \\
 & \left.+\left(k_{1}+k_{2}\right)b_{i}\dot{e}_{i}+\left(1+k_{1}k_{2}\right)b_{i}e_{i}+\hat{\nu}_{i}\right)\label{eq: Controller}
\end{align}
where $\hat{\nu}_{i}\in\mathbb{R}^{n}$ is designed as a Filippov
solution to
\begin{equation}
\dot{\hat{\nu}}_{i}=\left(k_{1}+\left(1+k_{1}k_{2}\right)k_{3}\right)\eta_{i}+k_{4}\text{sgn}\left(\dot{\eta}_{i}+k_{1}\eta_{i}\right),\label{eq: nuHatDot}
\end{equation}
where $k_{1},k_{2},k_{3},k_{4}\in\mathbb{R}_{>0}$ are user-defined
constants. To aid in the stability analysis, the graph interaction
matrix $\mathcal{H}\in\mathbb{R}^{nN\times nN}$ is defined as
\begin{equation}
\mathcal{H}\triangleq\left(L+B\right)\otimes I_{n},\label{eq:Augmented Interaction Matrix}
\end{equation}
where $B\triangleq\text{diag}\left\{ b_{1},\ldots,b_{N}\right\} \in\mathbb{R}^{N\times N}$.
Using (\ref{eq:Augmented Interaction Matrix}), the neighborhood position
error in (\ref{eq:relative position error 2}) is expressed in an
ensemble representation as
\begin{equation}
\eta=\mathcal{H}e,\label{eq:Neighborhood Position Ensemble}
\end{equation}
where $\eta\triangleq\left(\eta_{i}\right)_{i\in\mathcal{V}}\in\mathbb{R}^{nN}$
and $e\triangleq\left(e_{i}\right)_{i\in\mathcal{V}}\in\mathbb{R}^{nN}$.
Based on the subsequent stability analysis, define the filtered tracking
error $r_{1}\in\mathbb{R}^{nN}$ as
\begin{equation}
r_{1}\triangleq\dot{e}+k_{1}e.\label{eq: Filtered Tracking Error Ensemble}
\end{equation}
Using (\ref{eq:Neighborhood Position Ensemble}), (\ref{eq: Filtered Tracking Error Ensemble})
is expressed as
\begin{equation}
\mathcal{H}r_{1}=\dot{\eta}+k_{1}\eta.\label{eq: Filtered Tracking Error Ensemble.}
\end{equation}
Similarly, define the auxiliary tracking error $r_{2}\in\mathbb{R}^{nN}$
as
\begin{equation}
r_{2}\triangleq\dot{r}_{1}+k_{2}r_{1}+e.\label{eq:auxiliary tracking error ensemble}
\end{equation}
Using (\ref{eq: Filtered Tracking Error Ensemble.}), (\ref{eq:auxiliary tracking error ensemble})
is expressed as
\begin{equation}
\mathcal{H}r_{2}=\ddot{\eta}+\left(k_{1}+k_{2}\right)\dot{\eta}+\left(1+k_{1}k_{2}\right)\eta.\label{eq: Auxiliary Tracking Error ensemble}
\end{equation}
Using (\ref{eq: nuHatDot}), (\ref{eq: Filtered Tracking Error Ensemble.}),
and (\ref{eq: Auxiliary Tracking Error ensemble}), the time-derivative
of (\ref{eq: Controller}) is expressed in an ensemble representation
as{\footnotesize{}
\begin{align}
\dot{u} & =g^{+}\left(\mathcal{H}\left(r_{1}+k_{3}r_{2}\right)+k_{4}\text{sgn}\left(\mathcal{H}r_{1}\right)+\mathcal{B}\left(k_{1}\ddot{e}+k_{2}\dot{r}_{1}+\dot{e}\right)\right),\label{eq:ensemble controller}
\end{align}
}where $u\triangleq\left(u_{i}\right)_{i\in\mathcal{V}}\in\mathbb{R}^{N\sum_{i\in\mathcal{V}}m_{i}}$,
$g^{+}\triangleq\text{diag}\left(g_{1}^{+},\ldots,g_{N}^{+}\right)\in\mathbb{R}^{N\left(\sum_{i\in\mathcal{V}}m_{i}\times n\right)}$,
and $\mathcal{B}\triangleq B\otimes I_{n}\in\mathbb{R}^{nN}$. Substituting
(\ref{eq:ensemble controller}) into the time-derivative of (\ref{eq:auxiliary tracking error ensemble})
yields{\small{}
\begin{align}
\dot{r}_{2} & =h_{B}+\widetilde{h}-k_{4}\text{sgn}\left(\mathcal{H}r_{1}\right)-\left(k_{3}\mathcal{H}+\left(k_{1}+k_{2}\right)\left(\mathcal{B}-I_{nN}\right)\right)r_{2}\nonumber \\
 & -\left(\mathcal{H}+\left(1-2k_{1}^{2}-k_{1}k_{2}\right)\left(\mathcal{B}-I_{nN}\right)\right)r_{1}\nonumber \\
 & +\left(k_{1}\left(2-k_{1}^{2}\right)+k_{2}\right)\left(\mathcal{B}-I_{nN}\right)e,\label{eq:auxiliary tracking error ensemble dynamics}
\end{align}
}where $h_{B}\triangleq\left(\dot{f}_{0}\right.\left(q_{0},\dot{q}_{0},t\right)-\dot{f}_{i}\left(q_{0},\dot{q}_{0},t\right)-\dot{d}_{i}\left(t\right)-\dot{g}_{i}\left(q_{0},\dot{q}_{0},t\right)g_{i}^{+}\left(q_{0},\dot{q}_{0},t\right)\left.\left(\ddot{q}_{0}-f_{i}\left(q_{0},\dot{q}_{0},t\right)-d_{i}\left(t\right)\right)\right)_{i\in\mathcal{V}}\in\mathbb{R}^{nN}$
and $\widetilde{h}\triangleq\left(\dot{f}_{i}\left(q_{0},\dot{q}_{0},t\right)\right.-\dot{f}_{i}\left(q_{i},\dot{q}_{i},t\right)+\dot{g}_{i}\left(q_{0},\dot{q}_{0},t\right)g_{i}^{+}\left(q_{0},\dot{q}_{0},t\right)\left(\ddot{q}_{0}-f_{i}\left(q_{0},\dot{q}_{0},t\right)-d_{i}\left(t\right)\right)-\dot{g}_{i}\left(q_{i},\dot{q}_{i},t\right)g_{i}^{+}\left(q_{i},\dot{q}_{i},t\right)\left.\left(\ddot{q}_{i}-f_{i}\left(q_{i},\dot{q}_{i},t\right)-d_{i}\left(t\right)\right)\right)_{i\in\mathcal{V}}\in\mathbb{R}^{nN}$.
\begin{assumption}
\label{graph} \cite[Assumption 3]{Klotz.Kan.ea2015} The graph $G$
is connected with at least one $b_{i}>0$, for all $i\in\mathcal{V}$.
\end{assumption}
\begin{lem}
\label{HPD} \cite[Remark 1.2]{Ren.Cao2013} Following Assumption
\ref{graph}, $\mathcal{H}$ is positive definite.
\end{lem}
Define the concatenated state vector $z:\mathbb{R}_{\geq0}\to\mathbb{R}^{3nN}$
as $z\triangleq\left[\begin{array}{ccc}
e^{\top} & r_{1}^{\top} & r_{2}^{\top}\end{array}\right]^{\top}$. Using (\ref{eq: Filtered Tracking Error Ensemble}), (\ref{eq:auxiliary tracking error ensemble}),
and (\ref{eq:auxiliary tracking error ensemble dynamics}), the closed-loop
error system is expressed as{\small{}
\begin{alignat}{1}
\dot{z} & =\left[\begin{array}{c}
r_{1}-k_{1}e\\
r_{2}-k_{2}r_{1}-e\\
\left(\begin{array}{c}
h_{B}+\widetilde{h}-k_{4}\text{sgn}\left(\mathcal{H}r_{1}\right)\\
-\left(k_{3}\mathcal{H}+\left(k_{1}+k_{2}\right)\left(\mathcal{B}-I_{nN}\right)\right)r_{2}\\
-\left(\mathcal{H}+\left(1-2k_{1}^{2}-k_{1}k_{2}\right)\left(\mathcal{B}-I_{nN}\right)\right)r_{1}\\
+\left(k_{1}\left(2-k_{1}^{2}\right)+k_{2}\right)\left(\mathcal{B}-I_{nN}\right)e,
\end{array}\right)
\end{array}\right].\label{eq:Closed-Loop Error System}
\end{alignat}
}The following section provides a Lyapunov-based stability analysis
to provide exponential tracking error convergence guarantees with
the developed controller. Some supporting lemmas are first presented
to facilitate the subsequent analysis.
\begin{lem}
\label{Hb bounds}There exist known constants $\chi_{1},\chi_{2}\in\mathbb{R}_{>0}$
such that $\left\Vert h_{B}\right\Vert \leq\chi_{1}$ and $\left\Vert \dot{h}_{B}\right\Vert \leq\chi_{2}$.\footnote{If uncertainty bounds $\chi_{1},\chi_{2}$ are unknown, adaptive RISE
methods (e.g., \cite{Bidikli.Tatlicioglu.ea2014}) can estimate related
bounds online. This typically involves a trade-off, as such adaptive
approaches often yield asymptotic stability or ultimate boundedness,
contrasting with the exponential stability targeted herein, which
generally remains an open challenge for adaptive RISE.}
\end{lem}
\begin{IEEEproof}
By Assumption \ref{targetbounds}, $q_{0}$ is defined over a compact
set $\mathcal{Q}\subset\mathbb{R}^{n}$, where $\mathcal{Q}\triangleq\left\{ \iota\in\mathbb{R}^{n}:\left\Vert \iota\right\Vert \leq\overline{q}_{0}\right\} $.
Using the definition of $h_{B}$ and the triangle inequality yields
$\left\Vert h_{B}\right\Vert \leq N\left\Vert \frac{{\rm d}}{{\rm d}t}\right.f_{0}\left(q_{0},\dot{q}_{0},t\right)-\frac{{\rm d}}{{\rm d}t}f_{i}\left(q_{0},\dot{q}_{0},t\right)-\frac{{\rm d}}{{\rm d}t}d_{i}\left(t\right)-\frac{{\rm d}}{{\rm d}t}\left(g_{i}\left(q_{0},\dot{q}_{0},t\right)\right)g_{i}^{+}\left(q_{0},\dot{q}_{0},t\right)\left(f_{i}\left(q_{0},\dot{q}_{0},t\right)\right.+d_{i}\left(t\right)-\left.\left.\ddot{q}_{0}\right)\right\Vert $.
Applying Assumption \ref{targetbounds}, the chain-rule, the Cauchy-Schwarz
inequality, and the triangle inequality yields $\left\Vert h_{B}\right\Vert \leq N\left(\overline{\dot{q}}_{0}\right.\left(\left\Vert \frac{\partial f_{0}}{\partial q_{0}}\right\Vert +\left\Vert \frac{\partial f_{i}}{\partial q_{0}}\right\Vert \right)+\overline{\ddot{q}}_{0}\left(\left\Vert \frac{\partial f_{0}}{\partial\dot{q}_{0}}\right\Vert +\left\Vert \frac{\partial f_{i}}{\partial\dot{q}_{0}}\right\Vert \right)+\left\Vert \frac{\partial f_{0}}{\partial t}\right\Vert +\left\Vert \frac{\partial f_{i}}{\partial t}\right\Vert +\overline{\dot{d}}_{i}+\left(\left\Vert \frac{\partial g_{i}}{\partial q_{0}}\right\Vert \overline{\dot{q}}_{0}\right.+\left\Vert \frac{\partial g_{i}}{\partial\dot{q}_{0}}\right\Vert \overline{\ddot{q}}_{0}+\left.\left\Vert \frac{\partial g_{i}}{\partial t}\right\Vert \right)\left\Vert g_{i}^{+}\left(q_{0},\dot{q}_{0},t\right)\right\Vert \left(\left\Vert f_{i}\left(q_{0},\dot{q}_{0},t\right)\right\Vert \right.+\overline{d}_{i}+\left.\left.\overline{\ddot{q}}_{0}\right)\right)$.
Since $q_{0}$ is defined over the compact set $\mathcal{Q}$, applying
the mean value theorem gives $\left\Vert h_{B}\right\Vert \leq\chi_{1}$.

Using the definition of $h_{B}$ and the triangle inequality yields
$\left\Vert \dot{h}_{B}\right\Vert \leq N\left\Vert \frac{{\rm d}^{2}}{{\rm d}t^{2}}\right.f_{0}\left(q_{0},\dot{q}_{0},t\right)-\frac{{\rm d}^{2}}{{\rm d}t^{2}}f_{i}\left(q_{0},\dot{q}_{0},t\right)-\frac{{\rm d}^{2}}{{\rm d}t^{2}}d_{i}\left(t\right)-\frac{{\rm d}^{2}}{{\rm d}t^{2}}\left(g_{i}\left(q_{0},\dot{q}_{0},t\right)\right)g_{i}^{+}\left(q_{0},\dot{q}_{0},t\right)\left(f_{i}\left(q_{0},\dot{q}_{0},t\right)\right.+d_{i}\left(t\right)-\left.\left.\ddot{q}_{0}\right)\right\Vert $.
Applying Assumption \ref{targetbounds}, the chain-rule, the product
rule, the Cauchy-Schwarz inequality, and the triangle inequality yields$\left\Vert \dot{h}_{B}\right\Vert \leq N\left(\overline{\dot{q}}_{0}\left(\left\Vert \frac{{\rm d}}{{\rm d}t}\left(\frac{\partial f_{0}}{\partial q_{0}}\right)\right\Vert +\left\Vert \frac{{\rm d}}{{\rm d}t}\left(\frac{\partial f_{i}}{\partial q_{0}}\right)\right\Vert \right)\right.+\overline{\ddot{q}}_{0}\left(\left\Vert \frac{\partial f_{0}}{\partial q_{0}}\right\Vert +\left\Vert \frac{{\rm d}}{{\rm d}t}\left(\frac{\partial f_{0}}{\partial\dot{q}_{0}}\right)\right\Vert +\left\Vert \frac{\partial f_{i}}{\partial q_{0}}\right\Vert +\left\Vert \frac{{\rm d}}{{\rm d}t}\left(\frac{\partial f_{i}}{\partial\dot{q}_{0}}\right)\right\Vert \right)+\overline{\dddot{q}}_{0}\left(\left\Vert \frac{\partial f_{0}}{\partial\dot{q}_{0}}\right\Vert +\left\Vert \frac{\partial f_{i}}{\partial\dot{q}_{0}}\right\Vert \right)+\left\Vert \frac{{\rm d}}{{\rm d}t}\left(\frac{\partial f_{0}}{\partial t}\right)\right\Vert +\left\Vert \frac{{\rm d}}{{\rm d}t}\left(\frac{\partial f_{i}}{\partial t}\right)\right\Vert +\overline{\ddot{d}}_{i}+\left(\left\Vert \frac{{\rm d}}{{\rm d}t}\left(\frac{\partial g_{i}}{\partial q_{0}}\right)\right\Vert \right.\overline{\dot{q}}_{0}+\overline{\ddot{q}}_{0}\left(\left\Vert \frac{\partial g_{i}}{\partial q_{0}}\right\Vert +\left\Vert \frac{{\rm d}}{{\rm d}t}\left(\frac{\partial g_{i}}{\partial\dot{q}_{0}}\right)\right\Vert \right)+\left\Vert \frac{\partial g_{i}}{\partial\dot{q}_{0}}\right\Vert \left.\overline{\dddot{q}}_{0}+\left\Vert \frac{{\rm d}}{{\rm d}t}\left(\frac{\partial g_{i}}{\partial t}\right)\right\Vert \right)\left\Vert g_{i}^{+}(q_{0},\dot{q}_{0},t)\right\Vert \left(\left\Vert f_{i}(q_{0},\dot{q}_{0},t)\right\Vert \right.+\overline{d}_{i}+\left.\left.\overline{\ddot{q}}_{0}\right)\right)$.
Since $q_{0}$ is defined over the compact set $\mathcal{Q}$, applying
the mean value theorem gives $\left\Vert \dot{h}_{B}\right\Vert \leq\chi_{2}$.
\end{IEEEproof}
\begin{lem}
\label{rho function}\cite[Lemma 5]{Kamalapurkar.Rosenfeld.ea2014}
There exists a strictly increasing function $\rho:\left[0,\infty\right)\to\left[0,\infty\right)$
such that $\left\Vert \widetilde{h}\right\Vert \leq\rho\left(\left\Vert z\right\Vert \right)\left\Vert z\right\Vert $.
\end{lem}

\section{Stability Analysis}

To establish exponential stability for the closed-loop error system
in (\ref{eq:Closed-Loop Error System}) a $P$-function is introduced.
This function is used to develop a strict Lyapunov function and is
designed to be non-negative under specific gain conditions. The $P$-function
$P:\mathcal{I}\to\mathbb{R}_{\geq0}$ is designed as{\footnotesize{}
\begin{align}
P & \triangleq k_{4}\left\Vert \mathcal{H}r_{1}\right\Vert _{1}-r_{1}^{\top}\mathcal{H}^{\top}h_{B}+\mathrm{e}^{-\lambda_{P}t}*r_{1}^{\top}\mathcal{H}^{\top}\dot{h}_{B}\nonumber \\
 & +\mathrm{e}^{-\lambda_{P}t}*\left(k_{4}\left(k_{2}-\lambda_{P}\right)\left\Vert \mathcal{H}r_{1}\right\Vert _{1}-\left(k_{2}-\lambda_{P}\right)r_{1}^{\top}\mathcal{H}^{\top}h_{B}\right),\label{eq: P function}
\end{align}
}where $\lambda_{P}\in\mathbb{R}_{>0}$ is a user-defined constant,
and the symbol `$*$' denotes the convolutional integral, defined
for any given function $\alpha\left(t\right):\mathcal{I}\to\mathbb{R}$
as $\mathrm{e}^{-\lambda_{P}t}*\alpha\left(t\right)=\int_{t_{0}}^{t}\mathrm{e}^{-\lambda_{P}\left(t-\tau\right)}\alpha\left(t\right){\rm d}\tau$.
Using Leibniz's rule, the time derivative of a convolution integral
satisfies $\frac{{\rm d}}{{\rm d}t}\left(\mathrm{e}^{-\lambda_{P}t}*\alpha\right)=\alpha\left(t\right)-\lambda_{P}\int_{t_{0}}^{t}\mathrm{e}^{-\lambda_{P}\left(t-\tau\right)}\alpha\left(\tau\right){\rm d}t$
which simplifies to $\frac{{\rm d}}{{\rm d}t}\left(\mathrm{e}^{-\lambda_{P}t}*\alpha\right)=\alpha\left(t\right)-\lambda_{P}\mathrm{e}^{-\lambda_{P}t}*\alpha\left(t\right)$.
Using (\ref{eq: P function}) and the definition of the convolutional
integral yields
\begin{equation}
P\left(t_{0}\right)=k_{4}\left\Vert \mathcal{H}r_{1}\left(t_{0}\right)\right\Vert _{1}-r_{1}^{\top}\left(t_{0}\right)\mathcal{H}^{\top}h_{B}\left(t_{0}\right).\label{eq:Pt_0}
\end{equation}

Since $t\mapsto\mathcal{H}r_{1}\left(t\right)$ is absolutely continuous
and $\left\Vert \cdot\right\Vert _{1}$ is globally Lipschitz, the
mapping $t\mapsto\left\Vert r_{1}\left(t\right)\right\Vert _{1}$
is differentiable almost everywhere. By the chain rule from \cite[Theorem 2.2]{Shevitz.Paden1994},
the derivative of $\left\Vert \mathcal{H}r_{1}\right\Vert _{1}$ is
given almost everywhere by $\frac{{\rm d}}{{\rm d}t}\left\Vert \mathcal{H}r_{1}\right\Vert _{1}=\frac{{\rm d}}{{\rm d}t}\left\Vert \dot{\eta}+k_{1}\eta\right\Vert _{1}\stackrel{{\rm a.e.}}{\in}\left(\ddot{\eta}+k_{1}\dot{\eta}\right)^{\top}K\left[\text{sgn}\right]\left(\dot{\eta}+k_{1}\eta\right)$.
By the definition of the $L_{1}$-norm, $\left\Vert \dot{\eta}+k_{1}\eta\right\Vert _{1}=\left(\dot{\eta}+k_{1}\eta\right)^{\top}K\left[\text{sgn}\right]\left(\dot{\eta}+k_{1}\eta\right)$.
Using the definitions of the chain rule and the $L_{1}$-norm, taking
the time-derivative of (\ref{eq: P function}), using Leibniz's rule,
and substituting (\ref{eq: Filtered Tracking Error Ensemble.}), (\ref{eq: Auxiliary Tracking Error ensemble}),
and (\ref{eq: P function}) into the resulting expression yields that
$t\mapsto P\left(t\right)$ satisfies the differential inclusion

\begin{alignat}{1}
\dot{P} & \stackrel{{\rm a.e.}}{\in}-\lambda_{P}P-r_{2}^{\top}\mathcal{H}^{\top}\left(h_{B}-k_{4}K\left[\text{sgn}\right]\left(\mathcal{H}r_{1}\right)\right).\label{eq:PDot}
\end{alignat}
To facilitate the inclusion of the $P$-function into the subsequent
Lyapunov function candidate, $P$ must be designed to be non-negative
under certain sufficient gain conditions.
\begin{prop}
\label{prop:Ppositive} For the $P$-function defined as in (\ref{eq: P function}),
if $k_{4}>\chi_{1}+\frac{\chi_{2}}{k_{2}-\lambda_{P}}$ with $\lambda_{P}\in\left(0,k_{2}\right)$,
then $P\left(t\right)\geq0$, for all $t\in\mathcal{I}$.
\end{prop}
\begin{IEEEproof}
By Lemma \ref{Hb bounds}, the $k_{4}\left\Vert \mathcal{H}r_{1}\right\Vert _{1}-r_{1}^{\top}\mathcal{H}^{\top}h_{B}$
term in (\ref{eq: P function}) satisfies $k_{4}\left\Vert \mathcal{H}r_{1}\right\Vert _{1}-r_{1}^{\top}\mathcal{H}^{\top}h_{B}\geq\left(k_{4}-\chi_{1}\right)\left\Vert \mathcal{H}r_{1}\right\Vert _{1}\geq0$
by the Cauchy-Schwarz inequality and the condition $k_{4}>\chi_{1}$.
The convolution integrand in (\ref{eq: P function}) is bounded as
$r_{1}^{\top}\mathcal{H}^{\top}\dot{h}_{B}+\left(k_{2}-\lambda_{P}\right)\left(k_{4}\left\Vert \mathcal{H}r_{1}\right\Vert _{1}-r_{1}^{\top}\mathcal{H}^{\top}h_{B}\right)\geq\left\Vert \mathcal{H}r_{1}\right\Vert _{1}\left(\left(k_{2}-\lambda_{P}\right)\left(k_{4}-\chi_{1}\right)-\chi_{2}\right)$.
This expression is non-negative when $\left(k_{2}-\lambda_{P}\right)\left(k_{4}-\chi_{1}\right)\geq\chi_{2}$
which is equivalent to $k_{4}\geq\chi_{1}+\frac{\chi_{2}}{k_{2}-\lambda_{P}}$.
Since $k_{4}>\chi_{1}+\frac{\chi_{2}}{k_{2}-\lambda_{P}}$ by assumption,
the convolution integrand is positive, making the entire convolution
positive. Therefore, $P\left(t\right)\geq0$, for all $t\in\mathcal{I}$.
\end{IEEEproof}
To state the main results, the following definitions are introduced.
Let $W:\mathbb{R}^{3nN}\to\mathbb{R}_{\geq0}$ be defined as
\begin{align}
W\left(\sigma\right) & \triangleq\underline{\lambda}_{Q}^{-\frac{1}{2}}\sqrt{\overline{\lambda}_{Q}\left\Vert \sigma\right\Vert ^{2}+2\left(k_{4}+\chi_{1}\right)\left\Vert \mathcal{H}\right\Vert _{1}\left\Vert \sigma\right\Vert _{1}},\label{eq:W}
\end{align}
for all $\sigma\in\mathbb{R}^{3nN}$, where $Q\triangleq\text{diag}\left(I_{2nN},\mathcal{H}\right)\in\mathbb{R}^{3nN\times3nN}$
and $\chi_{1}$ is defined as in Lemma \ref{Hb bounds}. Additionally,
let $k_{\min}\in\mathbb{R}$ be a constant gain defined as $k_{\min}\triangleq\min\left\{ k_{1}\right.-\frac{1}{2},k_{2}-\frac{1}{2},2\left(k_{1}+k_{2}\right)\left(\underline{\lambda}_{\mathcal{H}}\underline{\lambda}_{\left(\mathcal{B}-I_{nN}\right)}+k_{3}\underline{\lambda}_{\mathcal{H}}^{2}\right)-\left(\overline{\lambda}_{\left(I_{nN}-\mathcal{H}^{2}\right)}+\left(1+2k_{1}^{2}+k_{1}k_{2}\right)\overline{\lambda}_{\mathcal{H}}\overline{\lambda}_{\left(\mathcal{B}-I_{nN}\right)}\right)^{2}-\left(k_{1}\left(2-k_{1}^{2}\right)+k_{2}\right)^{2}\overline{\lambda}_{\mathcal{H}}^{2}\left.\overline{\lambda}_{\left(\mathcal{B}-I_{nN}\right)}^{2}\right\} $,
$\lambda_{V}\in\mathbb{R}_{>0}$ be the desired rate of convergence,
and the set of stabilizing initial conditions $\mathcal{S}\subset\mathbb{R}^{3nN}$
be defined as
\begin{equation}
\mathcal{S}\triangleq\left\{ \sigma\in\mathbb{R}^{3nN}:\rho\left(W\left(\sigma\right)\right)\leq\frac{k_{\min}-\lambda_{V}}{\overline{\lambda}_{\mathcal{H}}}\right\} .\label{eq:basin of attraction}
\end{equation}
Let $\xi:\mathcal{I}\to\mathbb{R}^{3nN+1}$ be defined as $\xi\left(t\right)=\left[\begin{array}{cc}
z^{\top}\left(t\right) & P\left(t\right)\end{array}\right]^{\top}$, and $\psi:\mathbb{R}^{3nN+1}\times\mathbb{R}_{\geq0}\rightrightarrows\mathbb{R}^{3nN+1}$
denote the set-valued map{\small{}
\begin{equation}
\psi\left(\xi,t\right)\triangleq\left[\begin{array}{c}
r_{1}-k_{1}e\\
r_{2}-k_{2}r_{1}-e\\
\left(\begin{array}{c}
h_{B}+\widetilde{h}-k_{4}K\left[\text{sgn}\right]\left(\mathcal{H}r_{1}\right)\\
-\left(k_{1}+k_{2}\right)\left(\mathcal{B}-I_{nN}\right)r_{2}\\
-\left(1-2k_{1}^{2}-k_{1}k_{2}\right)\left(\mathcal{B}-I_{nN}\right)r_{1}\\
-\mathcal{H}r_{1}-k_{3}\mathcal{H}r_{2}\\
+\left(k_{1}\left(2-k_{1}^{2}\right)+k_{2}\right)\left(\mathcal{B}-I_{nN}\right)e
\end{array}\right)\\
\left(\begin{array}{c}
-\lambda_{P}P-r_{2}^{\top}\mathcal{H}^{\top}h_{B}\\
+k_{4}r_{2}^{\top}\mathcal{H}^{\top}K\left[\text{sgn}\right]\left(\mathcal{H}r_{1}\right)
\end{array}\right)
\end{array}\right].\label{eq: set-valued map}
\end{equation}
}Then using (\ref{eq:Closed-Loop Error System}) and (\ref{eq:PDot}),
it follows that the trajectories $t\mapsto\xi\left(t\right)$ satisfy
the differential inclusion $\dot{\xi}\stackrel{{\rm a.e.}}{\in}\psi\left(\xi,t\right)$.
\begin{thm}
All solutions to (\ref{eq:Closed-Loop Error System}) with $z\left(t_{0}\right)\in\text{int}\left(\mathcal{S}\right)$
satisfy $\left\Vert z\left(t\right)\right\Vert \leq W\left(z\left(t_{0}\right)\right){\rm e}^{-\frac{2\lambda_{V}}{\overline{\lambda}_{Q}}\left(t-t_{0}\right)}$,
for all $t\in\left[t_{0},\infty\right)$, provided that the sufficient
control gains $k_{1},k_{2},k_{3}$ are selected to satisfy $k_{1}>\frac{1}{2}$,
$k_{2}>\frac{1}{2}$, $k_{3}>\left(\overline{\lambda}_{\left(I_{nN}-\mathcal{H}^{2}\right)}+\overline{\lambda}_{\mathcal{H}}\overline{\lambda}_{\left(\mathcal{B}-I_{nN}\right)}\right)^{2}+2\overline{\lambda}_{\mathcal{H}}^{2}\overline{\lambda}_{\left(\mathcal{B}-I_{nN}\right)}^{2}$,
$k_{4}$ is selected to satisfy Proposition \ref{prop:Ppositive},
and Assumptions \ref{targetbounds} and \ref{graph} hold.
\end{thm}
\begin{IEEEproof}
Consider the function $V:\mathbb{R}^{3nN+1}\times\mathbb{R}_{\geq0}\to\mathbb{R}_{\geq0}$
defined as
\begin{equation}
V\left(\xi\right)\triangleq\frac{1}{2}z^{\top}Qz+P.\label{eq: Lyapunov Function Candidate}
\end{equation}
For the subsequent analysis, consider all arbitrary trajectories $t\mapsto\xi\left(t\right)$
satisfying (\ref{eq:Pt_0}) and $z\left(t_{0}\right)\in\text{int}\left(\mathcal{S}\right)$.
Since Proposition \ref{prop:Ppositive} holds, it follows that these
trajectories satisfy $P\left(t\right)\geq0$ for all $t\in\mathcal{I}$
which implies that $V\left(\xi\left(t\right)\right)>0$ for all $t\in\mathcal{I}$.
Invoking the Rayleigh quotient theorem and using (\ref{eq: Lyapunov Function Candidate})
yields
\begin{equation}
\frac{\underline{\lambda}_{Q}}{2}\left\Vert z\right\Vert ^{2}+P\leq V\left(\xi\right).\label{eq:Rayleigh quotient-1}
\end{equation}

Based on the chain rule for differential inclusions in \cite[Theorem 2.2]{Shevitz.Paden1994},
the derivative of $t\mapsto V\left(\xi\left(t\right)\right)$ exists
almost everywhere and is a solution to $\dot{V}\left(\xi\right)\stackrel{{\rm a.e.}}{\in}\dot{\widetilde{V}}\left(\xi\right)$,
where the set $\dot{\widetilde{V}}\left(\xi\right)$ is defined as
$\dot{\widetilde{V}}\left(\xi\right)\triangleq\underset{\zeta\in\partial V\left(\xi\right)}{\bigcap}\zeta^{\top}\psi\left(\xi,t\right)$.
Since $V\left(\xi\right)$ is continuously differentiable for all
$\xi\in\mathbb{R}^{3nN+1}$, Clarke's gradient reduces to the singleton
$\left\{ \nabla V\left(\xi\right)\right\} =\left\{ \left[\begin{array}{cccc}
e^{\top} & r_{1}^{\top} & r_{2}^{\top}\mathcal{H} & 1\end{array}\right]^{\top}\right\} $. Thus, $\dot{\widetilde{V}}\left(\xi\right)=\underset{\zeta\in\partial V\left(\xi\right)}{\bigcap}\zeta^{\top}\psi\left(\xi,t\right)=\left(\nabla V\left(\xi\right)\right)^{\top}\psi\left(\xi,t\right)$.
Evaluating $\dot{\widetilde{V}}\left(\xi\right)$ yields
\begin{align}
\dot{\widetilde{V}}\left(\xi\right) & =-k_{1}e^{\top}e-k_{2}r_{1}^{\top}r_{1}-\lambda_{P}P+r_{2}^{\top}\mathcal{H}\widetilde{h}\nonumber \\
 & -r_{2}^{\top}\left(\left(k_{1}+k_{2}\right)\mathcal{H}\left(\mathcal{B}-I_{nN}\right)+k_{3}\mathcal{H}^{2}\right)r_{2}\nonumber \\
 & +r_{2}^{\top}\left(I_{nN}-\mathcal{H}^{2}-\left(1-2k_{1}^{2}-k_{1}k_{2}\right)\mathcal{H}\left(\mathcal{B}-I_{nN}\right)\right)r_{1}\nonumber \\
 & +\left(k_{1}\left(2-k_{1}^{2}\right)+k_{2}\right)r_{2}^{\top}\mathcal{H}\left(\mathcal{B}-I_{nN}\right)e\nonumber \\
 & +k_{4}r_{2}^{\top}\mathcal{H}^{\top}\left(K\left[\text{sgn}\right]\left(\mathcal{H}r_{1}\right)-K\left[\text{sgn}\right]\left(\mathcal{H}r_{1}\right)\right).\label{eq:VDot-1}
\end{align}
By \cite[Footnote 2]{Klotz.Kan.ea2015}, the right-hand side of (\ref{eq:VDot-1})
is continuous almost everywhere. Specifically, it is discontinuous
only on a set of times with Lebesgue measure zero, where $K\left[\text{sgn}\right]\left(\mathcal{H}r_{1}\right)-K\left[\text{sgn}\right]\left(\mathcal{H}r_{1}\right)\neq0$.
Since this set is Lebesgue negligible, $K\left[\text{sgn}\right]\left(\mathcal{H}r_{1}\right)\stackrel{{\rm a.e.}}{=}\left\{ \text{sgn}\left(\mathcal{H}r_{1}\right)\right\} $.
Furthermore, from Lemma \ref{rho function}, $\left\Vert \widetilde{h}\right\Vert \leq\rho\left(\left\Vert z\right\Vert \right)\left\Vert z\right\Vert $.
Applying the Cauchy-Schwarz inequality, the triangle inequality, and
Young's inequality yields that (\ref{eq:VDot-1}) is upper bounded
as

\begin{align}
\dot{\widetilde{V}}\left(\xi\right) & \stackrel{{\rm a.e.}}{\leq}-\left(k_{1}-\frac{1}{2}\right)\left\Vert e\right\Vert ^{2}-\left(k_{2}-\frac{1}{2}\right)\left\Vert r_{1}\right\Vert ^{2}\nonumber \\
 & -\left(k_{1}+k_{2}\right)\left(\underline{\lambda}_{\mathcal{H}}\underline{\lambda}_{\left(\mathcal{B}-I_{nN}\right)}+k_{3}\underline{\lambda}_{\mathcal{H}}^{2}\right)\left\Vert r_{2}\right\Vert ^{2}\nonumber \\
 & +\left(\overline{\lambda}_{\left(I_{nN}-\mathcal{H}^{2}\right)}+\left(1+2k_{1}^{2}+k_{1}k_{2}\right)\overline{\lambda}_{\mathcal{H}}\overline{\lambda}_{\left(\mathcal{B}-I_{nN}\right)}\right)^{2}\left\Vert r_{2}\right\Vert ^{2}\nonumber \\
 & +\frac{1}{2}\left(k_{1}\left(2-k_{1}^{2}\right)+k_{2}\right)^{2}\overline{\lambda}_{\mathcal{H}}^{2}\overline{\lambda}_{\left(\mathcal{B}-I_{nN}\right)}^{2}\left\Vert r_{2}\right\Vert ^{2}\nonumber \\
 & +\overline{\lambda}_{\mathcal{H}}\rho\left(\left\Vert z\right\Vert \right)\left\Vert z\right\Vert ^{2}-\lambda_{P}P.\label{eq:VDot-4}
\end{align}
Since $k_{1},k_{2}$, and $k_{3}$ are selected to satisfy $k_{1}>\frac{1}{2}$,
$k_{2}>\frac{1}{2}$, and $k_{3}>\left(\overline{\lambda}_{\left(I_{nN}-\mathcal{H}^{2}\right)}+\overline{\lambda}_{\mathcal{H}}\overline{\lambda}_{\left(\mathcal{B}-I_{nN}\right)}\right)^{2}+2\overline{\lambda}_{\mathcal{H}}^{2}\overline{\lambda}_{\left(\mathcal{B}-I_{nN}\right)}^{2}$,
it follows that $k_{\min}>0$. Consequently, using the definition
of $\left\Vert z\right\Vert $ yields that (\ref{eq:VDot-4}) is upper-bounded
as
\begin{align}
\dot{\widetilde{V}}\left(\xi\right) & \stackrel{{\rm a.e.}}{\leq}-\left(k_{\min}-\overline{\lambda}_{\mathcal{H}}\rho\left(\left\Vert z\right\Vert \right)\right)\left\Vert z\right\Vert ^{2}-\lambda_{P}P.\label{eq:VDot-6}
\end{align}
Since $z\left(t_{0}\right)\in\mathcal{S}$, it follows from (\ref{eq:basin of attraction})
that $k_{\min}>\text{\ensuremath{\lambda_{V}+\overline{\lambda}_{\mathcal{H}}\rho\left(W\left(z\left(t_{0}\right)\right)\right)}}$.
Therefore,{\footnotesize{}
\begin{align}
\dot{\widetilde{V}}\left(\xi\right) & \stackrel{{\rm a.e.}}{\leq}-\left(\lambda_{V}+\overline{\lambda}_{\mathcal{H}}\left(\rho\left(W\left(z\left(t_{0}\right)\right)\right)-\rho\left(\left\Vert z\right\Vert \right)\right)\right)\left\Vert z\right\Vert ^{2}-\lambda_{P}P.\label{Vdot}
\end{align}
}By the definition of $W$ in (\ref{eq:W}), it follows that $\left\Vert z\left(t_{0}\right)\right\Vert \leq W\left(z\left(t_{0}\right)\right)$.
Because the solution $t\mapsto z\left(t\right)$ is continuous, $z$
cannot instantaneously escape $\mathcal{S}$ at $t_{0}$. Therefore,
there exists a time interval $\mathcal{I}_{\mathcal{S}}$ satisfying
$\mathcal{I}_{\mathcal{S}}\subseteq\mathcal{I}$ such that $z\left(t\right)\in\mathcal{S}$
for all $t\in\mathcal{I}_{\mathcal{S}}$, implying $\left\Vert z\left(t\right)\right\Vert <W\left(z\left(t_{0}\right)\right)$.
Because $\rho$ is strictly increasing, $\rho\left(\left\Vert z\left(t\right)\right\Vert \right)<\rho\left(W\left(z\left(t_{0}\right)\right)\right)$
for all $t\in\mathcal{I}_{\mathcal{S}}$. Consequently, using (\ref{eq:Rayleigh quotient-1}),
selecting $\lambda_{P}>\lambda_{V}$ and recalling $\dot{V}\left(\xi\right)\stackrel{{\rm a.e.}}{\in}\dot{\widetilde{V}}\left(\xi\right)$
yields that
\begin{align}
\dot{V}\left(\xi\right) & \leq-\frac{2\lambda_{V}}{\overline{\lambda}_{Q}}V\left(\xi\right),\label{eq:vdot}
\end{align}
for all $t\in\mathcal{I}_{\mathcal{S}}$. Solving the differential
inequality in (\ref{eq:vdot}) and using (\ref{eq:Rayleigh quotient-1})
yields that $\left\Vert z\right\Vert $ is upper-bounded as
\begin{equation}
\left\Vert z\left(t\right)\right\Vert \leq\sqrt{2\underline{\lambda}_{Q}^{-1}V\left(\xi\left(t_{0}\right)\right)}{\rm e}^{-\frac{2\lambda_{V}}{\overline{\lambda}_{Q}}\left(t-t_{0}\right)},\label{eq:ztbound}
\end{equation}
for all $t\in\mathcal{I}_{\mathcal{S}}$. Using (\ref{eq:Pt_0}) yields
that $V\left(\xi\left(t_{0}\right)\right)=\frac{1}{2}\overline{\lambda}_{Q}\left\Vert z\left(t_{0}\right)\right\Vert ^{2}+k_{4}\left\Vert \mathcal{H}r_{1}\left(t_{0}\right)\right\Vert _{1}-r_{1}^{\top}\left(t_{0}\right)\mathcal{H}^{\top}h_{B}\left(t_{0}\right)$.
Since $h_{B}$ is bounded by Lemma \ref{Hb bounds}, using the Cauchy-Schwarz
inequality and the fact that $\left\Vert r_{1}\left(t_{0}\right)\right\Vert \leq\left\Vert r_{1}\left(t_{0}\right)\right\Vert _{1}\leq\left\Vert z\left(t_{0}\right)\right\Vert _{1}$
yields
\begin{align}
\sqrt{2\underline{\lambda}_{Q}^{-1}V\left(\xi\left(t_{0}\right)\right)} & \leq W\left(z\left(t_{0}\right)\right),\label{eq:WBound}
\end{align}
 where $W$ is defined in (\ref{eq:W}). Finally, applying (\ref{eq:WBound})
to (\ref{eq:ztbound}) yields $\left\Vert z\left(t\right)\right\Vert \leq W\left(z\left(t_{0}\right)\right){\rm e}^{-\frac{2\lambda_{V}}{\overline{\lambda}_{Q}}\left(t-t_{0}\right)}$,
for all $t\in\mathcal{I}_{\mathcal{S}}$.

It remains to be shown that $\mathcal{I}_{\mathcal{S}}$ can be extended
to $\left[t_{0},\infty\right)$. Let $t\mapsto\xi\left(t\right)$
be a maximal solution to the differential inclusion $\dot{\xi}\stackrel{{\rm a.e.}}{\in}\psi\left(\xi,t\right)$
with initial conditions satisfying (\ref{eq:Pt_0}) and $z\left(t_{0}\right)\in\text{int}\left(\mathcal{S}\right)$.
From the preceding analysis, $z\left(t\right)\in\text{int}\left(\mathcal{S}\right)$
for all $t\in\mathcal{I}_{\mathcal{S}}$. This implies that $\xi\left(t\right)\in\mathcal{D}\triangleq\left\{ \mathfrak{S}\in\mathbb{R}^{3nN+1}:\mathfrak{S}=\left[\begin{array}{cc}
\sigma^{\top} & \varsigma\end{array}\right]^{\top},\sigma\in\text{int}\left(\mathcal{S}\right),\varsigma\in\mathbb{R}_{\geq0}\right\} $. For any compact subinterval $\mathcal{J}\subseteq\mathcal{I}_{\mathcal{S}}$,
$\left\Vert z\left(t\right)\right\Vert \leq W\left(z\left(t_{0}\right)\right){\rm e}^{-\frac{2\lambda_{V}}{\overline{\lambda}_{Q}}\left(t-t_{0}\right)}$,
for all $t\in\mathcal{J}$. Since $W$ is a locally bounded function
and ${\rm e}^{-\frac{2\lambda_{V}}{\overline{\lambda}_{Q}}\left(t-t_{0}\right)}\leq1$
for all $t\geq t_{0}$, the mapping $t\mapsto z\left(t\right)$ is
uniformly bounded on $\mathcal{J}$. By definition, $\left(\xi,t\right)\mapsto\psi\left(\xi,t\right)$
is locally bounded when $\xi$ is bounded. Since $z\left(t\right)$
is bounded on $\mathcal{J}$ and $P\left(t\right)$ is non-negative,
the trajectory $t\mapsto\left(\xi\left(t\right),t\right)$ is precompact.
Furthermore, since the map $\left(\xi,t\right)\mapsto\psi\left(\xi,t\right)$
is locally bounded, it follows from \cite[Remark 3.4]{Kamalapurkar.Dixon.ea2020}
that $\bigcup_{t\in\mathcal{J}}\psi\left(\xi\left(t\right),t\right)$
is bounded for every compact subinterval $\mathcal{J}\subseteq\mathcal{I}_{\mathcal{S}}$.
Therefore, the conditions of \cite[Lemma 3.3]{Kamalapurkar.Dixon.ea2020}
are satisfied, guaranteeing that $t\mapsto\xi\left(t\right)$ is complete,
i.e., $\mathcal{I}=\left[t_{0},\infty\right)$. Thus, all trajectories
satisfying $z\left(t_{0}\right)\in\text{int}\left(\mathcal{S}\right)$
also satisfy $\left\Vert z\left(t\right)\right\Vert \leq W\left(z\left(t_{0}\right)\right){\rm e}^{-\frac{2\lambda_{V}}{\overline{\lambda}_{Q}}\left(t-t_{0}\right)}$,
for all $t\in\left[t_{0},\infty\right)$.

Recall that $k_{\min}>\text{\ensuremath{\lambda_{V}+\overline{\lambda}_{\mathcal{H}}\rho\left(W\left(z\left(t_{0}\right)\right)\right)}}$,
implying that the exponential stability result is semi-global \cite[Remark 2]{Pettersen2017},
as the set of stabilizing initial conditions in (\ref{eq:basin of attraction})
can be made arbitrarily large by appropriately adjusting $k_{\min}$
to encompass any $z\left(t_{0}\right)\in\mathbb{R}^{3nN}$.

Because $k_{4}$, $\chi_{1}$, and $\lambda_{V}$ are independent
of the initial time $t_{0}$ or initial condition $z\left(t_{0}\right)$,
the exponential convergence is uniform \cite{Loria.Panteley2002}.
Additionally, the convergence and boundedness of $\left\Vert z\right\Vert $
implies the convergence and boundedness of $\left\Vert e\right\Vert $,
$\left\Vert r_{1}\right\Vert $, and $\left\Vert r_{2}\right\Vert $.
Therefore, since $q_{0},\dot{q}_{0},\ddot{q}_{0}\in\mathcal{L}_{\infty}$
by Assumption \ref{targetbounds}, using (\ref{eq: tracking error}),
(\ref{eq:relative position error 2}) and (\ref{eq:Neighborhood Position Ensemble})-(\ref{eq: Auxiliary Tracking Error ensemble})
yields that $q_{i},\dot{q}_{i},\ddot{q}_{i}\in\mathcal{L}_{\infty}$,
for all $i\in\mathcal{V}$. Thus, $f_{i}\left(q_{i},\dot{q}_{i},t\right)$
and $g_{i}\left(q_{i},\dot{q}_{i},t\right)$ are bounded for all $i\in\mathcal{V}$.
Therefore, using (\ref{eq:Agent Dynamics}) yields that $u\in\mathcal{L}_{\infty}$.
\end{IEEEproof}

\section{Simulation}

\begin{figure}
\begin{centering}
\includegraphics[width=1\columnwidth]{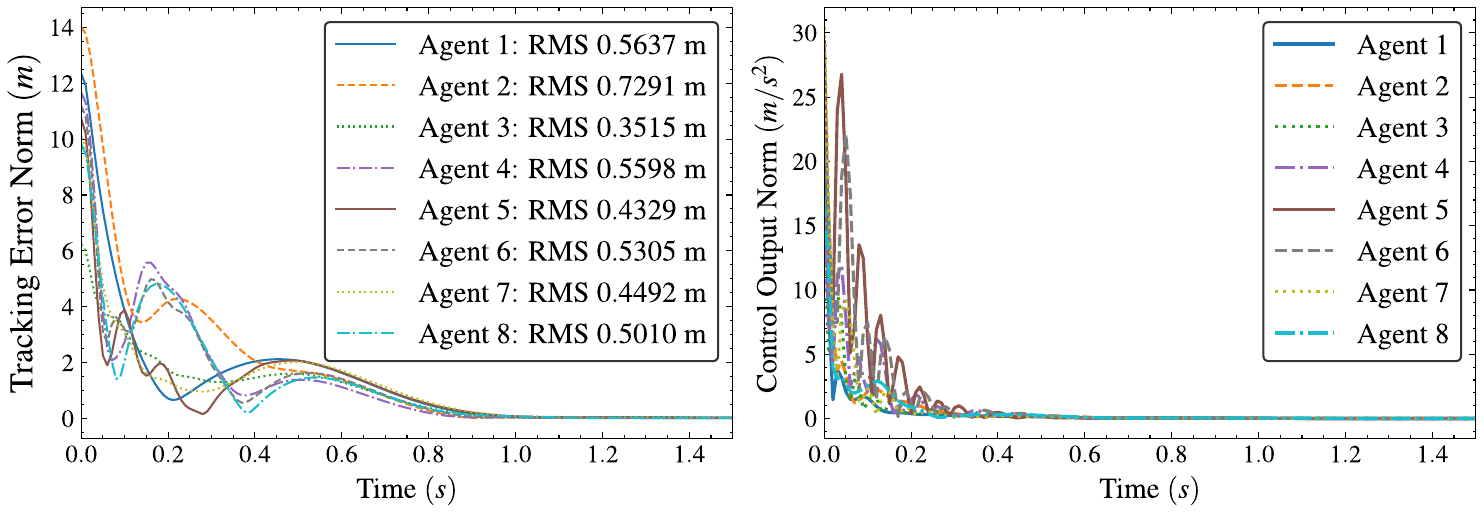}
\par\end{centering}
\caption{\label{fig:Tracking Error} Left: Tracking error norm for $t\in\left[0,1.5\right]$.
Right: Control effort norm for $t\in\left[0,1.5\right]$s.}
\end{figure}
Numerical simulations were performed to evaluate the proposed distributed
controller. The scenario involved $N=8$ agents tracking a dynamic
target agent, with inter-agent communication following a cycle graph
topology. The pinning matrix was $B=\text{diag}\left\{ 1,0,1,0,1,0,1,0\right\} $.
Agent initial positions were sampled from $U\left(-10,10\right)$
m per axis, with zero initial velocities. The target's initial position
was sampled from $U\left(-10,10\right)$ m with an initial velocity
of $\left[\begin{array}{ccc}
1 & -1 & 0.5\end{array}\right]^{\top}$ $m/s$. Each agent $i$ adhered to the dynamics in (\ref{eq:Agent Dynamics})
with $f_{i}=\left[c_{i,1}\left(y_{i}-z_{i}\right)\right.+c_{i,2}\tanh\left(\dot{x}_{i}t\right)c_{i,3}\left(z_{i}-x_{i}\right)+c_{i,4}\tanh\left(\dot{y}_{i}t\right)c_{i,5}\left(x_{i}-y_{i}\right)+\left.c_{i,6}\tanh\left(\dot{z}_{i}t\right)\right]^{\top}$,
$g_{i}=I_{3}-\text{diag}\left\{ c_{i,7}\cos\left(t\right),c_{i,8}\sin\left(t\right),c_{i,9}\cos\left(t\right)\sin\left(t\right)\right\} $,
and $d_{i}=\left[\begin{array}{ccc}
c_{i,10}\cos\left(t\right) & c_{i,11}\sin\left(t\right) & c_{i,12}\cos\left(t\right)\sin\left(t\right)\end{array}\right]^{\top}$ with parameters $c_{i,j}$ sampled from $U\left(-0.5,0.5\right)$.
The target's dynamics were $f_{0}=\left[\sin\left(x_{0}\right)\right.-\cos\left(y_{0}\dot{x}_{0}\right)\cos\left(z_{0}\dot{y}_{0}\right)-\sin\left(x_{0}\right)-\sin\left(y_{0}\dot{z}_{0}\right)-\left.\sin\left(z_{0}\right)\right]^{\top}$.

Relative position and velocity measurements were corrupted by additive
zero-mean Gaussian noise ($\sigma=0.001$ m and $\sigma=0.001$ m/s,
respectively). Furthermore, heterogeneous time delays, unique to each
agent, were introduced: communication delays for inter-agent information,
actuation delays for control signal application, and target sensing
delays for pinned agents were each randomly sampled for every agent
from $U\left(0.001,0.02\right)$ s. Control gains are $k_{1}=k_{2}=10$,
$k_{3}=25$, and $k_{4}=50$. The total simulation duration was 30
s. Fig. \ref{fig:Tracking Error} depicts the tracking error norms
and control effort norms, demonstrating robust target tracking despite
the uncertainties, noise, and varied delays, over a subset of the
30-second simulation.

\section{Conclusion}

A distributed RISE-based control framework is developed for multi-agent
target tracking that achieves exponential convergence while requiring
only local information exchange. The key innovation lies in the development
of a controller and Lyapunov-based stability analysis which results
from incorporating the graph interaction matrix directly into the
Lyapunov function, eliminating the need for 2-hop communication present
in previous approaches. The control development necessitated the development
of a novel $P$-function construction technique that accommodates
the interaction matrix while preserving the mathematical properties
required for stability guarantees. The nonsmooth stability analysis
established semi-global exponential convergence to the target agent
state despite the presence of bounded disturbances with bounded derivatives.\bibliographystyle{ieeetr}
\bibliography{Sources}

\end{document}